\documentclass[aps, superscriptaddress, showpacs, floatfix, twocolumn]{revtex4-1}
\usepackage{graphicx}
\usepackage{hyperref}
\usepackage{amsmath}
\usepackage{color}
\allowdisplaybreaks
\begin{document}
\title{Concurrence of topological electrons and magnons in the\\ Kagome ferromagnet CoCu$_3$(OH)$_6$Cl$_2$}
\author{Zhuoran He}
\thanks{These authors contributed equally to this work.}
\affiliation{Wuhan National High Magnetic Field Center $\&$ School of Physics, Huazhong University of Science and Technology, Wuhan 430074, China}
\author{Aiyun Luo}
\thanks{These authors contributed equally to this work.}
\affiliation{Wuhan National High Magnetic Field Center $\&$ School of Physics, Huazhong University of Science and Technology, Wuhan 430074, China}
\author{Biao Lian}
\affiliation{Princeton Center for Theoretical Science $\&$ Department of Physics, Princeton University, Princeton, New Jersey 08544, USA}
\author{Gang Xu}
\email[Electronic address: ]{gangxu@hust.edu.cn}
\affiliation{Wuhan National High Magnetic Field Center $\&$ School of Physics, Huazhong University of Science and Technology, Wuhan 430074, China}

\begin{abstract}
\vspace*{0\baselineskip}
Spin and charge are two interrelated properties of electrons. However, most of previous works on topological matter study the electronic and magnonic excitations separately. In this paper, by combining density functional theory calculations with the Schwinger boson method, we determine the topological electronic band structures and topological magnon spectrum simultaneously in the ferromagnetic ground state of the narrow-band-gap CoCu$_3$(OH)$_6$Cl$_2$, which is an ABC stacking Kagome lattice material with fractional occupancy on Cu sites. This material provides an ideal platform to study the interplay of different types of topological excitations. Our work also proposes a useful method to deal with correlated topological magnetic systems with narrow band gaps.
\end{abstract}

\maketitle
\textit{Introduction} --- The topological phases of matter have attracted intensive interest in the last decades. While the studies of topological states originally focus on electronic systems \cite{haldane1988model,kane2005z, bernevig2006quantum, qi2006topological, xiao2010berry, fu2011topological, zhang2009topological, yu2010quantized}, a route has been inspired towards the generalization of topological states to bosonic quasi-particles. For instance, topologically nontrivial bands have been characterized in photonic \cite{ozawa2019topological}, phononic \cite{stenull2016topological}, and magnonic \cite{zhang2013topological} systems. The topological magnon excitations are particularly interesting, which can be viewed as the magnetic counterpart of electronic systems in the vector space of spin waves. In the presence of strongly correlated electrons, the topological properties of the magnon bands and electron bands become intrinsically connected. Yet, their relationships and mutual influences are comparatively less studied and require more research efforts.

\begin{figure}[b]
\includegraphics[width=0.95\columnwidth]{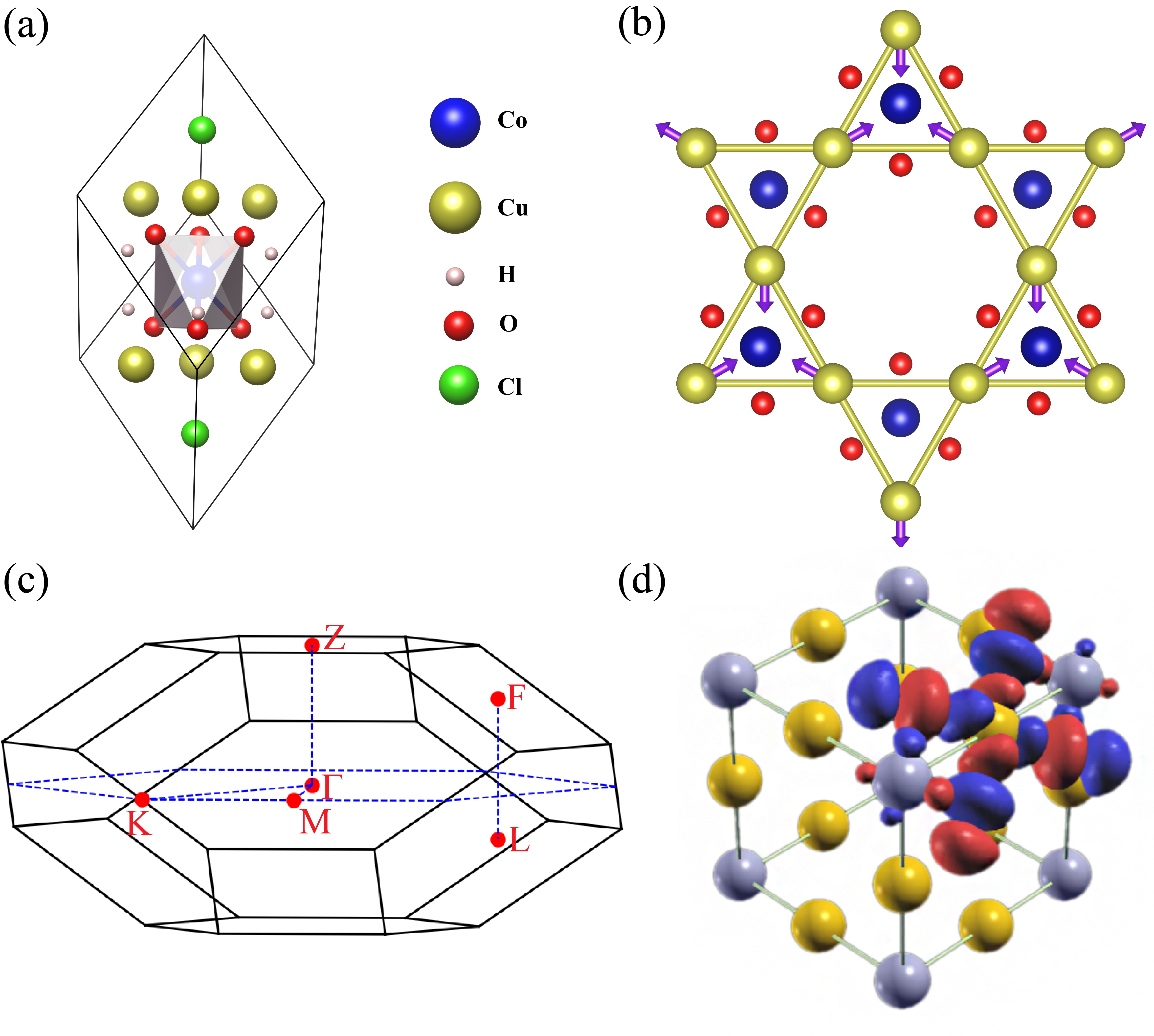}
\caption{(a) Rhombohedral primitive cell of CoCu$_3$(OH)$_6$Cl$_2$ with space group R-3m (No.~166). (b) Top view along the hexagonal $c$-axis to show the Cu Kagome layer. H and Cl ions are omitted. (c) Brillouin zone and high-symmetry path. (d) The hybridized Wannier orbital centered on a Cu site.
\label{fig:crystal}}
\end{figure}

The Kagome lattice provides an ideal platform to study the interplay of topological states and correlated electrons. The system has been known to yield flat bands and Dirac points \cite{guo2009topological}, which could further give rise to topological and Chern insulating phases by considering the spin-orbit coupling (SOC) and ferromagnetism \cite{xu2015intrinsic, ye2018massive, guterding2016prospect}. Furthermore, a class of Herbertsmithite-type Kagome materials exhibits spin liquids \cite{shores2005structurally}, viscous electron fluids \cite{di2019turbulent}, and $f$-wave superconductivity \cite{mazin2014theoretical}. In Kagome magnets, the bosonic magnon quasiparticles may have a Dirac-like spectrum or form topologically nontrivial insulating phases \cite{fransson2016magnon, li2017dirac, zhang2013topological}. Inelastic neutron scattering experiments have observed anomalous thermal Hall effects and protected chiral edge modes from the topological magnon band structures in insulating Kagome ferromagnets \cite{chisnell2015topological, onose2010observation}. Nonetheless, previous works on topological magnons mostly focus on the Heisenberg model \cite{li2017dirac, chisnell2015topological, onose2010observation}, in which electron charges are frozen and the whole system becomes a bosonic system composed of localized spins. However, this supposition does not apply to materials with narrow band gaps, where electron hoppings cannot be neglected, and the charge and spin degrees of freedom are both significant.

In this work, starting from density functional theory (DFT) calculations, we build a $t-J$ model to study the Kagome ferromagnet CoCu$_3$(OH)$_6$Cl$_2$. By using the Schwinger boson mean-field theory (SBMFT) \cite{PhysRevB.40.2610, PhysRevB.47.15192}, we take into account both the charge and the spin degrees of freedom and obtain topologically nontrivial fermionic and bosonic excitation spectra simultaneously. The fermionic spectrum well reproduces the DFT calculated electronic structures in energy dispersion and topological properties. Meanwhile, the bosonic spectrum reveals that CoCu$_3$(OH)$_6$Cl$_2$ is also a topological magnon system, in which the magnon thermal Hall effect can be expected. Our work demonstrates that the electron hopping and magnetic exchange coupling should be both considered in the narrow-band-gap Kagome magnet CoCu$_3$(OH)$_6$Cl$_2$, which provides a platform to study the interplay of topological electronic structures and topological magnon bands.

\begin{figure}
\includegraphics[width=\columnwidth]{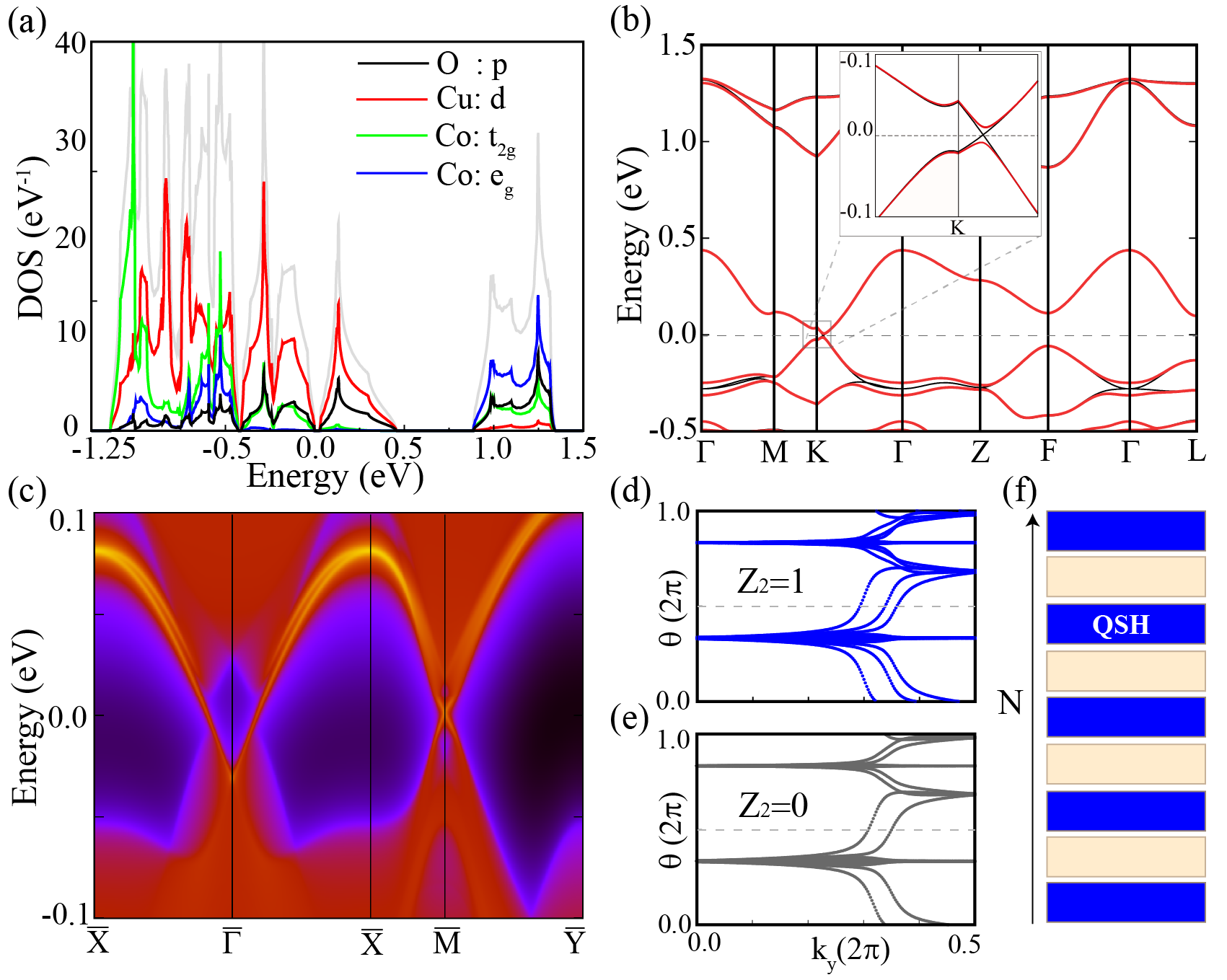}
\caption{(a) The projected DOS and (b) band structures with (colored) and without (black) SOC for the NM phase of CoCu$_3$(OH)$_6$Cl$_2$ at $U=J=0$. (c) The topological surface states of the weak TI phase in the (001) plane. (d) The Wilson loops of trilayer ($Z_2 = 1$) and (e) bilayer ($Z_2 = 0$) slabs. (f) Schematic of the oscillatory transitions between quantum spin Hall (QSH) and trivial insulators as a function of the total number of layers $N$.
\label{fig:NM}}
\end{figure}

\textit{DFT calculations} ---
The crystal structure of Herbertsmithite CoCu$_3$(OH)$_6$Cl$_2$ is illustrated in Figs.~\ref{fig:crystal}a \& \ref{fig:crystal}b, which crystallizes in the rhombohedral lattice with space group R-3m (No.~166). The Co ions are located at the centers of O$_6$-octahedra, which separate the ABC-stacking Kagome layers formed by the Cu ions. Our DFT calculations were performed by the Vienna ab initio simulation package (VASP) \cite{kresse1993ab, kresse1996efficient}. Perdew-Burke-Ernzerhof (PBE) type of the generalized gradient approximation + Hubbard $U$ (GGA+$U$) \cite{perdew1996generalized, anisimov1991band} was used as the exchange-correlation potential. We employed the experimental lattice parameters for all calculations. The cut-off energy for the wave function expansion was set to $400$~eV, and a $9\times9\times9$ k-mesh in the first Brillouin zone (BZ) was used for self-consistent calculations. The SOC was considered self-consistently. We then construct the maximally localized Wannier functions of the Cu $d_{x^2-y^2}\&d_{xy}$ antibonding states, hereafter referred to as $(d_{x^2-y^2}\&d_{xy})^*$, using  WANNIER90 \cite{mostofi2014updated} and calculate the topological properties using WannierTools \cite{wu2018wanniertools}.

We first calculate the nonmagnetic (NM) phase by GGA, and plot the density of states (DOS) with SOC and electronic band structures with/without SOC in Figs.~\ref{fig:NM}a \& \ref{fig:NM}b, respectively. By analyzing the projected DOS, it is clear that the states of the O, H and Cl ions are far from the Fermi level (0 eV), while all states between $-1.25\sim1.5$~eV are dominated by the $3d$-orbitals of Co and Cu ions. Since the Co ions are in an octahedral environment, the crystal-field effect will split the $3d$-orbitals into two manifolds -- the lower-energy $t_{2g}$ states between $-1.25\sim-0.5$~eV and the higher-energy $e_g$ states between $1.0\sim1.5$~eV, which suggests that the crystal field splitting is about 2.2 eV. The energy window of $-0.5\sim 0.5$~eV is dominated by the Cu $3d$-orbitals -- specifically the $(d_{x^2-y^2}\&d_{xy})^*$ states.

\begin{figure}
\includegraphics[width=\columnwidth]{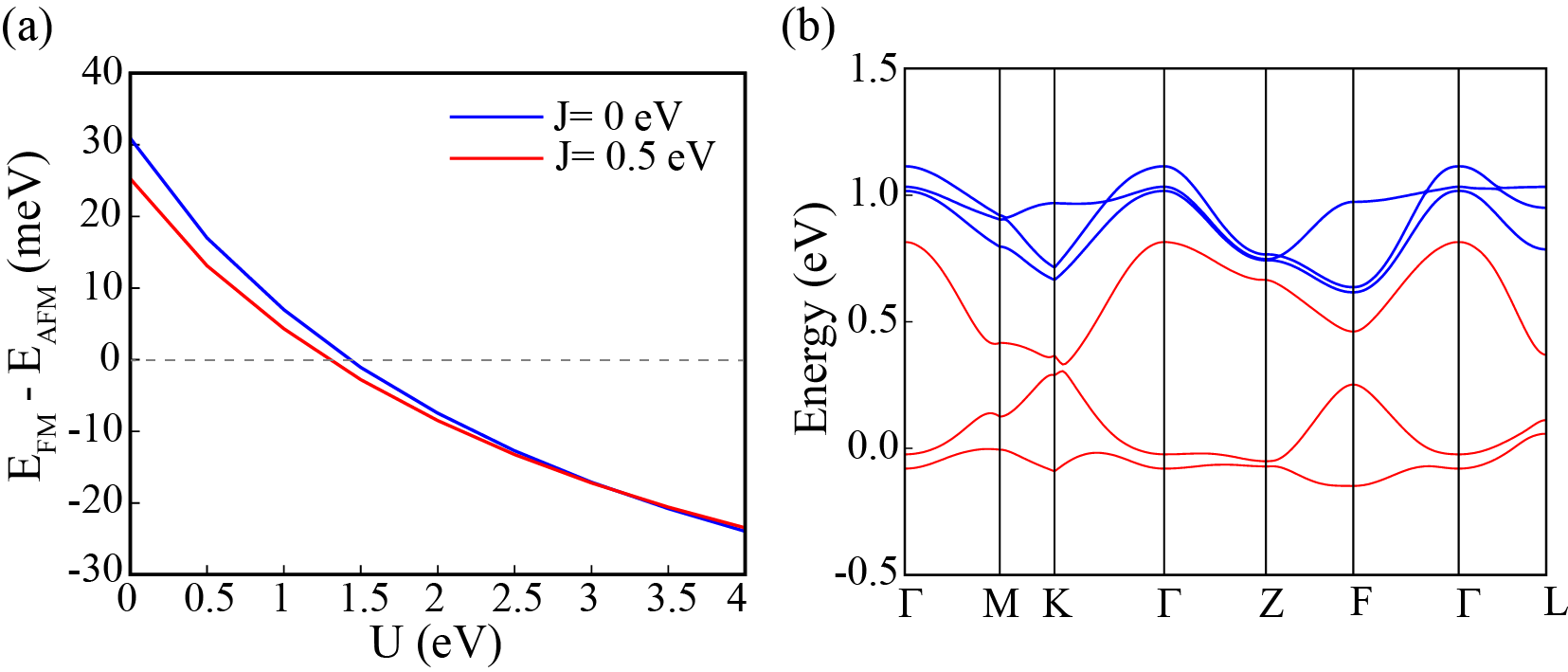}
\caption{(a) Energy difference of the FM and AFM phases versus $U$ and $J$ on Cu-$3d$. (b) Band structures for the FM ground state of CoCu$_3$(OH)$_6$Cl$_2$ with $U=4$~eV and $J=0.5$~eV, where the red bands are mainly from Cu-$3d$ orbitals and the blue bands are dominated by Co-$e_g$.
\label{fig:FM}}
\end{figure}

Our NM calculations suggest that the Co ion is trivalent. In the Herbertsmithite family $M$Cu$_3$(OH)$_6$Cl$_2$ ($M=\mathrm{Li}^+$, Na$^+$, Zn$^{2+}$, Mg$^{2+}$, Ga$^{3+}$, Sc$^{3+}$, etc.), the $M$ ions exhibit diverse valences, which give rise to different fillings of the Cu-$(d_{x^2-y^2}\&d_{xy})^*$ states and various topological properties \cite{guterding2016prospect, shores2005structurally, mazin2014theoretical, di2019turbulent}. We emphasize that the third ionization energy of Co is very close to that of Ga, and the octahedral environments in CoCu$_3$(OH)$_6$Cl$_2$ and GaCu$_3$(OH)$_6$Cl$_2$ are very similar. As a result, the Co ions tend to be trivalent, as opposed to the divalent Zn ions in ZnCu$_3$(OH)$_6$Cl$_2$. The Co$^{3+}$ ions will lead to the fractional valence of Cu$^{\frac{5}{3}+}$, which means that the filling number of $(d_{x^2-y^2}\&d_{xy})^*$ is 4/3 per Cu ion ($n=4/3$). Such filling makes the Dirac points cross the Fermi level exactly in the NM phase without the SOC, as shown by the black bands in Fig.~\ref{fig:NM}b. When the SOC is considered, a topologically nontrivial gap of about $20$~meV is opened, which gives rise to a weak topological insulator (TI) with the index $Z_2(\nu_0;\nu_1\nu_2\nu_3)=(0;111)$ as derived by the Fu-Kane formula \cite{fu2007topological}.

To demonstrate the bulk-boundary correspondence, we construct the Wannier functions of the Cu-$(d_{x^2-y^2}\&d_{xy})^*$ states to build a single-orbital tight-binding model in the Herbertsmithite Kagome materials. In Fig.~\ref{fig:crystal}d, we plot our Wannier function in real space and find that the first interlayer hopping is much weaker than the second interlayer hopping, and many qualitative features are correctly captured when only the latter is retained. This is the reason why we use the second interlayer hopping in our $t-J$ model for later discussions. By using the Green's function method, we investigate the surface states of the (001) plane as shown in Fig.~\ref{fig:NM}c. It is clear that there are two Dirac cones on the (001) plane, which is consistent with the conclusion that only an even number of surface Dirac cones can appear on crystal planes with Miller index $\vec{h} = \sum_{i}h_i b_i$ satisfying $h_i \neq \nu_i\;\!\mathrm{mod}\;\!2$ \cite{fu2007topological}. Moreover, $(\nu_1\nu_2\nu_3)=(111)$ demonstrates that this weak TI can be taken as 2D TI (QSH system) layers stacked along the (111) direction, i.e.,~$c$-axis of the hexagonal unit cell. Therefore, there will be oscillatory transitions between 2D QSH insulators and trivial insulators if one cleaves the weak TI into slabs along the $c$-axis. In Fig.~\ref{fig:NM}d, we plot the Wilson loop evolution for the slab with three layers, which confirms it fall into the 2D TI phase. In contrast, the Wilson loop of the slab with two layers shown in Fig.~\ref{fig:NM}e identifies it as a trivial insulator. Further calculations illustrate that the slabs with odd numbers of layers are 2D TIs, while the slabs with even numbers of layers are trivial, as depicted in Fig.~\ref{fig:NM}f. Such results agree with our expectations and are consistent with a similar crystal structure studied in Ref.~\cite{yan2012prediction}.

Considering that Co and Cu usually exhibit magnetism and correlation effects, we further calculate multiple magnetic structures by using the DFT + $U$ + SOC method. Interestingly, our calculations show that no magnetic moments were stabilized on Co$^{3+}$.  This is because the crystal field splitting (2.2~eV) is much larger than the typical Hund's coupling (0.5~eV) and the Zeeman splitting on Co$^{3+}$ (0.1~eV). As a result, the six Co-$3d$ electrons fully occupy the t$_{2g}$ bands, and no magnetic moment could be left on Co$^{3+}$. Similar properties also exist in other compounds with Co$^{3+}$ centered octahedra \cite{roth1964magnetic}. Based on the above analysis, we hereafter focus on the magnetism of Cu with ferromagnetic (FM) and frustrated antiferromagnetic (AFM) orders (as shown in Fig.~\ref{fig:crystal}b). We should note that both magnetic phases are much lower in energy than the NM phase in our calculations. The energy difference of FM and frustrated AFM is summarized in Fig.~\ref{fig:FM}a, which shows that the frustrated AFM phase is more stable than FM ordering if $U\leq1.5$~eV. For the typical range of $U$ of Cu ($U>3$~eV), FM ordering becomes more stable. The emergence of the FM ground state in CoCu$_3$(OH)$_6$Cl$_2$ is in distinct contrast to ZnCu$_3$(OH)$_6$Cl$_2$, in which strong AFM exchange interactions exist between the nearest neighboring Cu ions. Ferromagnetism is also suggested in GaCu$_3$(OH)$_6$Cl$_2\,$ \cite{mazin2014theoretical} and can be explained by the carriers on $(d_{x^2-y^2}\&d_{xy})^*$ away from half filling \cite{lin1987two}. The FM phase of CoCu$_3$(OH)$_6$Cl$_2$ has a magnetic moment of 2$\mu_B$ per primitive cell. All our results are insensitive to the Hund's coupling $J$ as shown in Fig.~\ref{fig:FM}a.

In Fig.~\ref{fig:FM}b, we plot the calculated FM results with $U=4$~eV to demonstrate the electronic band structures of the ground state of CoCu$_3$(OH)$_6$Cl$_2$. Obviously, strong Coulomb repulsion pushes the majority-spin bands far below the Fermi level, and only one electron occupies the minority-spin bands formed by $(d_{x^2-y^2}\&d_{xy})^*$ (marked red) due to the filling $n=4/3$. As shown in Fig.~\ref{fig:FM}b, the SOC will open topologically nontrivial gaps between the minority-spin bands, which are characterized by Chern numbers 1, 0 and $-1$ (from bottom to top) in a fixed $k_z$-plane. If the topologically nontrivial gap is large enough, one electron will totally occupy the $C=1$ band, and the system becomes a 3D layered Chern insulator \cite{xu2015intrinsic}, which provides the possibility to realize a quantum anomalous Hall (QAH) insulator by cleaving off a few layers \cite{deng2020quantum}.

\begin{figure}
\includegraphics[width=\columnwidth]{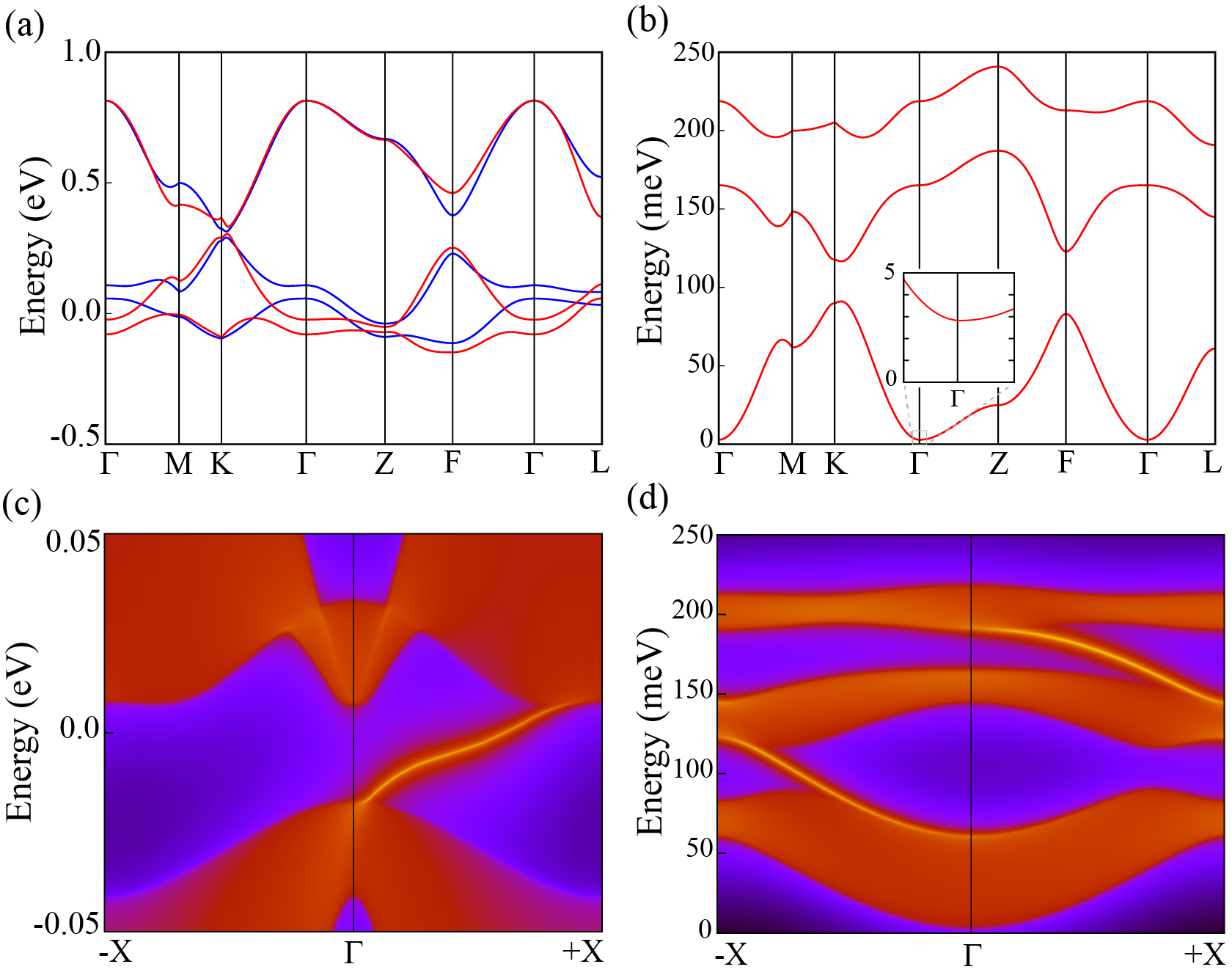}
\caption{(a) The minority-spin Cu-bands with SOC (blue) fitted by $t-J$ doublon bands (red), with $\tilde{t}_1=0.122\,(1+0.06i)$~eV, $\tilde{t}_2=0.3\,\mathrm{Re}\,\tilde{t}_1$, $\tilde{t}_3=0.1\,\mathrm{Re}\,\tilde{t}_1$. (b) Magnon bands with $J_1=-20$~meV and $J_2=-5$~meV. (c) The topological chiral edge modes of electrons and (d) topological chiral edge modes of magnons in the 2D reciprocal $k_z=0$ plane.
\label{fig:TB-band-fit}}
\end{figure}

\textit{$t-J$ model and magnon bands} --- To further study the interdependence of the low-energy electronic and magnonic excitations in CoCu$_3$(OH)$_6$Cl$_2$, we construct a Hubbard model on the 3D ABC stacking Kagome lattice for the Cu-$(d_{x2-y2}\&d_{xy})^*$ orbitals with the filling $n=4/3$:
\begin{align}
H_\mathrm{Hubbard} = \sum_{i\neq j}\sum_{\alpha\beta}t_{ij}^{\alpha\beta}c_{i\alpha}^\dagger c_{j\beta}+U\sum_i n_{i\uparrow} n_{i\downarrow},
\label{eq:H-Hubbard}
\end{align}
where $c_{i\alpha}^\dagger, c_{j\beta}$ creates/destructs an electron on sites $i,j$ with spin $\alpha,\beta=\;\uparrow,\downarrow$, $n_{i\alpha}=c_{i\alpha}^\dagger c_{i\alpha}$, $U$ is an on-site Coulomb repulsion and the hopping parameters $t_{ij}^{\alpha\beta}=t_{ij}\delta_{\alpha\beta}-i\vec{w}_{ij}\cdot\vec{\sigma}_{\alpha\beta}$ take into account the SOC, with $\vec{\sigma}$ being the Pauli vector and $\vec{w}_{ij}=-\vec{w}_{ji}$. According to previous studies \cite{PhysRevB.74.125106}, $U\sim 4$~eV is much greater than $ t\sim 0.2$~eV \cite{mazin2014theoretical} in Herbertsmithites. Thus, Eq.~\eqref{eq:H-Hubbard} can be reduced to a $t-J$ model $H_{t-J}= H_t+H_J$ given by
\begin{subequations}
\begin{align}
H_t&=\sum_{i\neq j}\sum_{\alpha\beta}\,t_{ij}^{\alpha\beta}\,Pc_{i\alpha}^\dagger c_{j\beta}P\,_{\!},\\
H_J&=\frac{1}{2}\sum_{i\neq j}J_{ij} \left(\vec{S}_i\cdot\vec{S}_j-\frac{1}{4}n_in_j\right)\!.
\end{align}
\end{subequations}
Here $\vec{S}_i=\frac{1}{2}\sum_{\alpha\beta}c_{i\alpha}^\dagger\vec{\sigma}_{\alpha\beta}c_{i\beta}$ and $n_i=\sum_{\alpha}c_{i\alpha}^\dagger c_{i\alpha}$ are the spin and charge operators. $J_{ij}$ is the Heisenberg exchange coupling. Since the Cu-$(d_{x2-y2}\&d_{xy})^*$ bands are over half-filling in Co-Herbertsmithite, the projection $P$ in Eq.~(2a) prohibits empty sites. Hence, the allowed site occupancies are the doublons and spinons in the slave-fermion representation $c_{i\sigma}^\dagger = b_{i\sigma}^\dagger f_i$ \cite{PhysRevB.61.3494, PhysRevB.73.214517,PhysRevLett.118.177201}, where the Schwinger bosons $b_{i\uparrow},b_{i\downarrow}$ describing spin and the fermionic doublons $f_i$ describing charge satisfy \cite{PhysRevB.40.2610}
\begin{align}
\sum_\sigma b_{i\sigma}^\dagger b_{i\sigma}=1-f_i^\dagger f_i.
\label{eq:spinon-chargeon-constraint}
\end{align}

According to our calculations, the main SOC effect comes from the in-plane nearest-neighbor hopping, which is an $S_z$-conserving term. We then neglect other SOC terms in $H_{t-J}$, which leads to $t_{ij}^{\alpha\beta}=t_{ij}^\alpha\delta_{\alpha\beta}$ (i.e.,~$\vec{w}_{ij}\parallel\hat{z}$). $H_{t-J}$ can now be recast into $S_z$-conserving mean-field forms of the auxiliary particles $b_{i\sigma}, f_i$:
\begin{subequations}
\begin{align}
H_{b\sigma} &= \sum_{ij}\left[(H_{b\sigma})_{ij}-\mu_{b\sigma}\delta_{ij}\right]b_{i\sigma}^\dagger b_{j\sigma},\\
H_f &= \sum_{ij}\left[(H_f)_{ij}-\mu_f\delta_{ij}\right]f_i^\dagger f_j,
\label{eq:Hb-sigma-mu}
\end{align}
\end{subequations}
where $\mu_{b\sigma}$ and $\mu_f$ are the chemical potentials satisfying $\mu_{b\uparrow}-\mu_{b\downarrow}=-\sum_jJ_{ij}\langle f_if_i^\dagger f_jf_j^\dagger\rangle\langle S_{jz}\rangle$ and the bosonic Hamiltonian is given by
\begin{align}
(H_{b\sigma})_{ij}=t_{ij}^\sigma\langle f_j^\dagger f_i\rangle+\frac{1}{2}J_{ij}\langle f_if_i^\dagger f_jf_j^\dagger\rangle\langle b_{j\bar{\sigma}}^\dagger b_{i\bar{\sigma}}\rangle.
\label{eq:Hb-real-space}
\end{align}
The fermionic Hamiltonian is given by
\begin{align}
(H_f)_{ij}=\sum_\sigma t_{ji}^\sigma\langle b_{j\sigma}^\dagger b_{i\sigma}\rangle+J_{ij}\langle A_{ij}^\dagger A_{ij}\rangle\langle f_j^\dagger f_i\rangle,
\label{eq:Hf-real-space}
\end{align}
with $A_{ij}^\dagger=\frac{1}{\sqrt{2}}\,_{\!}(b_{i\uparrow}^\dagger b_{j\downarrow}^\dagger-b_{i\downarrow}^\dagger b_{j\uparrow}^\dagger)$ creating a pair of spinons in the spin singlet state on bond $i-j$.
The expectation values $\langle A_{ij}^\dagger A_{ij}\rangle$ and $\langle f_if_i^\dagger f_jf_j^\dagger\rangle$ are evaluated within the Hartree-Fock approximation. We emphasize that, as shown in Eq.~\eqref{eq:Hf-real-space}, the fermionic bands can be affected by the Heisenberg couplings $J_{ij}$, which may even drive a topological phase transition in the fermionic bands.

By Fourier transform, Eq.~\eqref{eq:Hf-real-space} yields a $3\times3$ matrix $H_f(\vec{k})$ in momentum space, where the off-diagonal entries are $(H_f)_{ij}=2\tilde{t}_1\cos(\vec{k}\cdot\frac{\vec{a}_i-\vec{a}_j}{2})$ for $(H_f)_{21}$, $(H_f)_{32}$, and $(H_f)_{13}$. The diagonal entries $(H_f)_{ii}=2\tilde{t}_2\cos(\vec{k}\cdot\vec{a}_i)$\linebreak $+\,2\tilde{t}_3\,\Sigma_{j\neq i}'\cos[\vec{k}\cdot(\vec{a}_i-\vec{a}_j)]$. The effective hopping para-meters $\tilde{t}_{1-3}$ are obtained from the $t_{ji}^\sigma\,_{\!}\langle b_{j\sigma}^\dagger b_{i\sigma}\rangle$ term in Eq.~\eqref{eq:Hf-real-space} with $\sigma$ being the majority spin, while the minority-spin term and the $J_{ij}\langle A_{ij}^\dagger A_{ij}\rangle$ term both vanish in the FM phase \cite{PhysRevLett.117.227201}. The doublon bands with $\tilde{t}_1=0.122\,(1+0.06i)$~eV, $\tilde{t}_2=0.3\,\mathrm{Re}\,\tilde{t}_1$ and $\tilde{t}_3=0.1\,\mathrm{Re}\,\tilde{t}_1$ are plotted in Fig.~\ref{fig:TB-band-fit}a, which shows that our fermionic model Hamiltonian nicely reproduces both the energy dispersion and the topological properties of the electronic bands calculated by DFT.

Based on the filling $n=4/3$, the self-consistently determined doublon occupancy is $\langle f_i^\dagger f_i\rangle = 1/3$ in our $t-J$ model. Then according to Eq.~\eqref{eq:spinon-chargeon-constraint} and that the total magnetic moment is $2\mu_B$/p.c., we have $\langle b_{i\uparrow}^\dagger b_{i\uparrow}\rangle = 2/3$ and $\langle b_{i\downarrow}^\dagger b_{i\downarrow}\rangle=0$, confirming that the FM phase is realized in Co-Herbertsmithite. As a result, only the spin-up spinons exist in the bosonic ground state, in which all spinons condense at the $\Gamma$ point, and the magnonic excitation is given by the spin-down spinon bands, as discussed in Ref.~\cite{PhysRevLett.117.227201}. This is because creating a magnon in the FM phase is equivalent to destructing a spin-up spinon in the Bose condensate and recreating a spin-down spinon that carries the momentum of the magnon. The magnon excitation spectrum in Fig.~\ref{fig:TB-band-fit}b shows a mininum energy gap at the $\Gamma$ point $\Delta E_\Gamma=2.88$~meV, which is the lowest magnetic excitation energy. Our calculations show that $\Delta E_\Gamma$ is independent of the Heisenberg couplings $J_{ij}$, which is consistent with Ref.~\cite{liu2020spin}. Another important feature in Fig.~\ref{fig:TB-band-fit}b is that the three magnon bands are well-separated by the SOC-induced k-dependent Dzyaloshinskii-Moriya interactions \cite{PhysRevLett.120.197202}. We calculate the magnon Chern numbers in a fixed $k_z$ plane and obtain $C_{1-3}^{b\downarrow}=-1,0,1$ from bottom to top, which agree with previous studies on topological magnon insulators in Kagome ferromagnets \cite{zhang2013topological,chisnell2015topological}.

For any fixed $k_z$, the fermionic and bosonic Hamiltonians in Eqs.~\eqref{eq:Hb-real-space}--\eqref{eq:Hf-real-space} become 2D systems with nontrivial Chern numbers. We then use the Green's function method to calculate the edge states for $k_z=0$ and plot them in Figs.~\ref{fig:TB-band-fit}c \& \ref{fig:TB-band-fit}d. Fig.~\ref{fig:TB-band-fit}c shows the fermionic local density of states in the $\mathrm{X}$ direction. There is a chiral edge mode across the Fermi level propagating along $+\mathrm{X}$, which agrees with the Chern number $C=1$ of the occupied band from DFT calculations. In Fig.~\ref{fig:TB-band-fit}d, we illustrate the magnonic excitation spectrum of a semi-infinite quasi-1D sample along the $\mathrm{X}$ direction, which clearly shows two chiral edge modes in the bulk gaps of $90 \sim 115$~meV and $165 \sim 190$~meV. The edge modes in Fig.~\ref{fig:TB-band-fit}d have the same chirality and both propagate in the $-\mathrm{X}$ direction, which is a natural consequence of the Chern numbers $C_{1-3}^{b\downarrow}=-1,0,1$.  Since magnons can carry heat currents, the magnon Chern numbers are closely related to the thermal Hall conductivity in topological magnon systems \cite{katsura2010theory, mook2014magnon}. We expect that the magnon thermal Hall effect can be realized in FM Co-Herbertsmithite, which has been observed in other Kagome ferromagnets \cite{chisnell2015topological, onose2010observation}.

\textit{Conclusion} --- We have formulated a Schwinger boson \linebreak method for the spin-orbital coupled $t-J$ model constructed based on first-principles calculations, which gives the electronic and magnonic excitation spectra and their topological properties self-consistently. Our approach is advantageous for narrow-band-gap magnetic systems, where electrons do not freeze into fully localized spins as assumed in the Heisenberg model. We apply our method to the Co-Herbertsmithite and obtain 3D topological Chern bands of electrons and magnons \linebreak simultaneously. The present work demonstrates that the Schwinger boson method is suitable for narrow-band-gap correlated topological materials, where both the charge and spin contribute to the low-energy physics.

\textit{Acknowledgments} --- This work is supported by the National Key Research and Development Program of China (2018YFA0307000), and the National Natural Science Foundation of China (11874022). B. L. is supported by Princeton Center for Theoretical Science at Princeton University. B. L. and G. X. would like to thank the helpful support from Shou-Cheng Zhang at the early stage of this work.

\bibliography{refs}
\end{document}